\documentclass[structabstract]{aa}  
\usepackage{graphicx}
\usepackage{txfonts}
\begin{document}
   \title{A search for the near--infrared counterpart of the eclipsing millisecond X--ray pulsar Swift J1749.4--2807\thanks{Based on observations made with ESO Telescopes at the Paranal Observatory under programme ID 285.D-5030(A)}}


   \author{P. D'Avanzo
          \inst{1}
          \and
          S. Campana
	  \inst{1}
	  \and
	  T. Mu\~noz-Darias
	  \inst{1}
	  \and
	  T. Belloni
	  \inst{1}
	  \and
	  E. Bozzo
	  \inst{2}
	  \and
	  M. Falanga
	  \inst{3}
	  \and
	  L. Stella
	  \inst{4}
          }

   \institute{INAF, Osservatorio Astronomico di Brera, via E. Bianchi 46, I-23807 Merate (LC), Italy\\
              \email{paolo.davanzo@brera.inaf.it} 
         \and
           ISDC Data Center for Astrophysics of the University of Geneva, chemin d'\'Ecogia 16, 1290 Versoix, Switzerland
	   \and
	   International Space Science Institute (ISSI), Hallerstrasse 6, 3012 Bern, Switzerland  
	   \and
	   INAF, Osservatorio Astronomico di Roma, via Frascati 33, I-00040 Monteporzio Catone, Italy
             }

   \date{received; accepted}

 
  \abstract
   {Swift J1749.4--2807 is a transient accreting millisecond X-ray pulsars, the first that displayed X-ray eclipses. Therefore it holds a great potential for accurate mass measurements in a low
mass X-ray binary system.}
   {The determination of the companion star radial velocity would make it possible to fully resolve the system and to accurately measure the mass of the neutron star based on dynamical measurements. Unfortunately, no optical/NIR counterpart has been identified to date for this system, either in outburst or in quiescence.}
   {We performed a photometric study of the field of Swift J1749.4--2807 during quiescence in order to search for the presence of a variable counterpart. The source direction lies on the Galactic plane, making any search for its optical/NIR counterpart challenging. To minimize the effects of field crowding and interstellar extinction, we carried out our observations using the adaptive optics near-infrared imager NACO mounted at the ESO Very Large Telescope.}
   {From the analysis of public \it{Swift} {\rm X-ray data obtained during outburst, we derived the most precise ($1.6''$ radius) position for this source. Due to the extreme stellar crowding of the field, 41 sources are detected in our VLT images within the X--ray error circle, with some of them possibly showing variability consistent with the expectations.}}
   {{\rm We carried out the first deep imaging campaign devoted to the search of the quiescent NIR counterpart of Swift J1749.4--2807. Our results allow to provide constraints on the nature of the companion star of this system. Furthermore, they suggest that future phase-resolved NIR observations (performed with large aperture telescopes and adaptive optics) covering the full orbital period of the system are likely to identify the quiescent counterpart of Swift J1749.4--2807, through the measure of its orbital variability, opening the possibility of dynamical studies of this unique source.}}


   \maketitle
%

\section{Introduction}

Accreting millisecond X--ray pulsars (AMXPs) are a subclass of low-mass X--ray binary transients hosting an old, weakly magnetic ($10^8-10^9$ G) neutron star which accretes mass from a companion star and has been spun up to millisecond periods via transfer of angular momentum (for a review see Wijnands 2005). Since 1998, fourteen such systems have been discovered, with orbital periods in the range between 40 min and 19 hr and spin frequencies from 1.7 to 5.4 ms (Wijnands \& van der Klis 1998; Chakrabarty \& Morgan 1998; Markwardt et al. 2002; Galloway et al. 2002; Campana et al. 2003; Strohmayer et al. 2003; Galloway et al. 2005; Kaaret et al. 2006; Krimm et al. 2007; Casella et al. 2008; Altamirano et al. 2008; Altamirano et al. 2010; Papitto et al. 2010; \cite{Ma10}; \cite{Al11}; Papitto et al. 2011). Since they have been spun-up by accretion (see, e.g., Falanga et al. 2005), their initial mass should have increased by at least 0.1-0.2 $M_{\odot}$ compared to the canonical neutron star mass (e.g. Thorsett \& Chakrabarty 1999). AMXPs are therefore excellent systems to place tight constraints on the neutron star equation of state, a key result for fundamental physics.

The most direct way to determine the mass of an object in a binary system is through dynamical measurements. These sources are ideal candidates for such studies because by measuring the delays in the arrival times of the pulsations it is possible to derive with high precision the system's orbital period and the radial velocity of the neutron star ($K_1$). In addition, during the quiescent phase of transient X-ray binaries, the projected radial velocity of the companion star ($K_2$) can be measured through phase-resolved optical spectroscopy, when it is possible to study the light from the companion star without a sizeable contamination from the accretion disc (see, e.g., Hynes 2010). Alternatively, the same measurements can be carried out during outburst by measuring the Doppler shifts of narrow emission lines in the Bowen blend, originating from the irradiated face of the companion star (Steeghs \& Casares 2002). The only missing ingredient to fully resolve the system is its inclination ($i$) which is usually very hard to determine with precision. The presence of eclipses is very helpful in this respect (although these are observed only rarely). 

The AMXP Swift J1749.4--2807 was discovered in June 2006 by the {\it Swift} satellite when a type I burst from it was observed, which was emitted during a dim outburst (Wijnands et al. 2009; Campana 2009). The transient resumed activity in April 2010. During this outburst observations carried out with the {\it Rossi X-ray Timing Explorer} ({\it RossiXTE}) satellite led to the discovery of a neutron star spin period of 1.9 ms  and of an orbital period of 8.82 hr (\cite{Ma10}; \cite{Al11}). 
Short eclipses (with a duration of $\sim 2200$ s) were observed in the X--rays from observations with {\it RossiXTE}. In the {\it RossiXTE} and {\it Swift}-XRT light curves folded at the orbital period, marked flux decreases occur near orbital longitude 90 degrees (as measured from the ascending node), i.e. when the neutron star is expected to be behind the companion. {\it RossiXTE} observations cover two separate eclipse egresses, and one ingress (\cite{Ma10}). At least one X--ray eclipse was also observed in {\it Swift-}XRT data during the 2010 outburst (Ferrigno et al. 2011). During this eclipse, a residual flux is clearly seen in the {\it Swift}-XRT data (the flux reduces by a factor 3.3). This was ascribed to a dust scattering halo located along the line of sight to Swift J1749.4--2807 (Ferrigno et al. 2011). X--ray pulsations are not detected during these intervals, testifying to their genuine eclipse nature (\cite{Ma10}; \cite{Al11}; \cite{Fe11}). The detection of X--ray eclipses enables to constrain the system inclination in the $74.4^{\circ}-77.9^{\circ}$ range (depending on the assumed neutron star mass; \cite{Ma10}; \cite{Al11}).

Swift J1749.4--2807 is the only known AMXP showing eclipses and therefore would be a very promising low mass x--ray binary system for dynamical studies aimed at precise mass measurements. As discussed above, through the determination of $K_2$ it will be possible to precisely determine the mass of the neutron star with dynamical measurements. To this end, it is mandatory to identify its optical/NIR counterpart. Unfortunately, the source lies in the Galactic plane, in the direction of an extremely crowded and exinct region, making any search for its optical/NIR counterpart challenging. During the 2006 and 2010 outbursts no optical/NIR counterpart was detected (Kubanek et al. 2006; Khamitov et al. 2006; Blustin et al. 2006; Melandri et al. 2006; Yang et al. 2010). Furthermore, no X-ray counterpart could be detected with a 1.1 ks observation carried out with the {\it Chandra} satellite during the fading phase of the 2010 outburst (Chakrabarty, Jonker \& Markwardt 2010).

We present here the results of the first observational campaign devoted to the identification of the NIR quiescent counterpart of Swift J1749.4--2807. In order to minimize the effects of field crowding and interstellar extinction, we carried out our observations using the adaptive optics near-infrared imager NACO mounted at the ESO Very large Telescope.

\section{Observations and results}

\begin{figure*}
   \centering
   \includegraphics[angle=0,width=\textwidth]{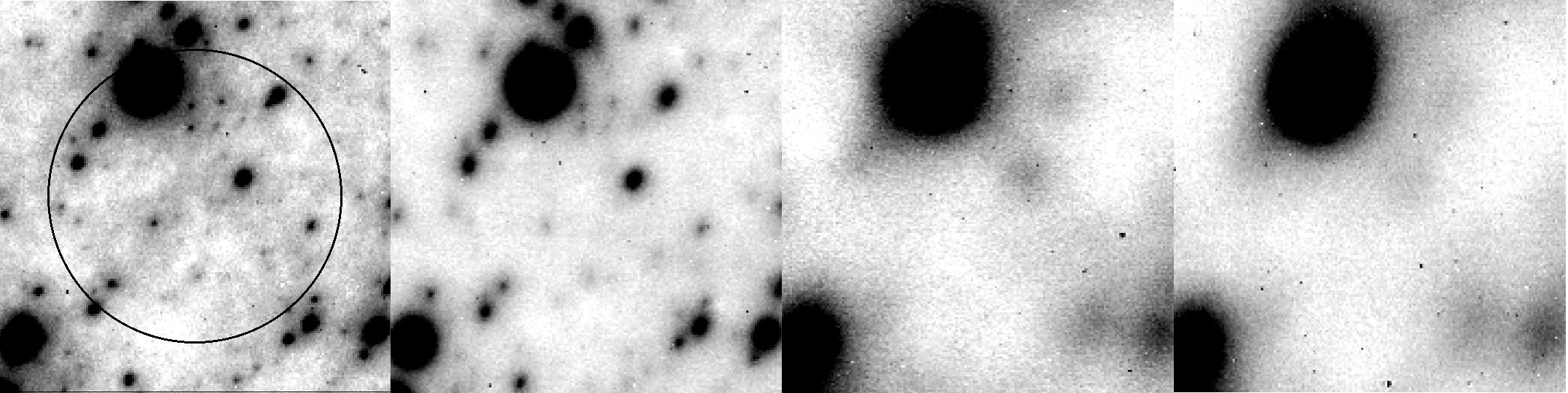}
      \caption{{\it From left to right}. The field of Swift J1749.4--2807 observed in the $H-$band on 2010 Aug 30, Aug 31, Sept 01, Sept 05. A clear worsening of resolution and flux limit is visible in the last two epochs (see Sect. 2 for details). The black circle represents the X--ray error circle (1.6'' radius, $90\%$ c.l.). North is up and East is left, image size is $5.4'' \times 5.4''$.
	    }
       \label{fig:fc1}
   \end{figure*}

We observed the field of Swift J1749.4--2807 with our approved DDT program ID 285.D-5030(A) at the ESO Very large Telescope (VLT). Observations were carried out in service mode on 2010 Aug 30th, 31st, Sept 1st and 5th with NAos COnica (NACO), the adaptive optics (AO) NIR imager and spectrometer mounted at
the VLT UT4 telescope. We used the S27 camera which has a pixel size of $0.027''$ and a field of view of $28'' \times 28''$. As a reference for the AO correction we used
the GSC-2 star S222131225301 (V=14.7), located $18''$ away from our target. The visual dichroic element and wavefront sensor were used. The observation log is presented in Table~\ref{tab:log}. 
Image reduction was carried out using the NACO pipeline data reduction, part of the ECLIPSE\footnote{http://www.eso.org/projects/aot/eclipse/} package. The images were dark subtracted, corrected for flat-field, bad and hot pixels and sky subtracted. In order to increase the signal-to-noise ratio, we averaged the images taken on each night. Unfortunately, the last two epochs of observations (Sept 1st and Sept 5th) were affected by natural seeing $> 1''$, with a resulting poorer resolution and shallower flux limit with respect to the others and were not included in our analysis (Fig.~\ref{fig:fc1}). Therefore, all the analysis and results presented in the following refer to the observing epochs Aug 30 and Aug 31. 
Astrometry was carried out by using the USNOB1.0\footnote{http://www.nofs.navy.mil/data/fchpix/} and the 2MASS\footnote{http://www.ipac.caltech.edu/2mass/} catalogues as reference with a resulting accuracy of $0.2''$. Point Spread Function (PSF)-photometry was made with the ESO-MIDAS\footnote{http://www.eso.org/projects/esomidas/} DAOPHOT task for all the objects in the field. 
The photometric calibration was done against standard stars. In order to minimize any systematic effect, we performed differential photometry with respect to a selection of local isolated and non-saturated reference stars. 

\begin{table*}
   \centering
\caption{Observation log.}
\begin{tabular}{cccccc}
\hline
UT mid observation &  Exposure		    &  PSF     &   Instrument  & Filter & Orbital phase \\
  (YYYmmdd)      &  (s)			    & (arcsec) &	       &        &               \\ 
\hline
20100830.0474(3) &  $7 \times 2 \times 60$ s  & $0.10''$  & VLT/NACO  & $H$    &   0.747(9)   \\
20100831.0533(3) &  $7 \times 2 \times 60$ s  & $0.17''$  & VLT/NACO  & $H$    &   0.485(9)   \\
20100901.0632(3) &  $7 \times 2 \times 60$ s  & $0.62''$  & VLT/NACO  & $H$    &   0.234(9)   \\
20100905.1124(3) &  $7 \times 2 \times 60$ s  & $0.71''$  & VLT/NACO  & $H$    &   0.256(9)   \\
\hline
\end{tabular}
\label{tab:log}
\end{table*}

We analysed the X--ray {\it Swift}-XRT data obtained during the 2006 and 2010 outbursts. We derived an enhanced position for the 2006 and 2010 outbursts, respectively, using the Leicester University interface\footnote{http://www.swift.ac.uk/user\_objects/}. We then combined the two enhanced positions to derive the following coordinates: RA(J2000)=17:49:31.83, Dec 
(J2000)=--28:08:04.7 with an uncertainty radius of $1.6''$ (90\% c.l.). This is the most precise position for this source obtained so far. As shown in Fig. 1, in spite of its relatively small size, the error circle is quite crowded, being the direction located towards the Galactic plane. Future {\it Chandra} X--ray observations would significantly reduce the uncertainty in the source position.

The VLT observing epochs were chosen so as to have a phase difference of $0.25\,P_{\rm orb}$ (based on the precise ephemeris of Altamirano et al. 2011). On Aug 30.05 UT the source was observed at phase $\sim 0.75$ (ascending node) while on Aug 31.05 it was at phase $\sim 0.5$ (inferior conjunction, i.e. when the observer sees the side of the companion star that faces the compact object). Under the assumption that the companion star is subject to irradiation from the compact object, as observed in all the compact ($P_{\rm orb} < 4.3$ hr) AMXPs for which a well-sampled optical light curve was obtained (D'Avanzo et al. 2009 and references therein), an increase in luminosity of a few tenths of magnitude is expected between the two epochs, making the identification of the NIR counterpart of the system easier. On the other hand, the 8.82 hr orbital period of Swift J1749.4--2807 implies a larger orbital separation than that of compact AMXPs, possibly reducing the effects of irradiation. In this case, a ``classical'' ellipsoidal modulation is expected for the optical/NIR light curve (as observed for the $P_{\rm orb} \sim 19$ hr AMXP Aql X-1; Shahbaz et al. 1998; Welsh et al. 2000) resulting in a fading between the two epochs.

\subsection{Candidate NIR counterparts of Swift J1749.4--2807}

\begin{figure}
   \centering
   \includegraphics[angle=-90,width=\columnwidth]{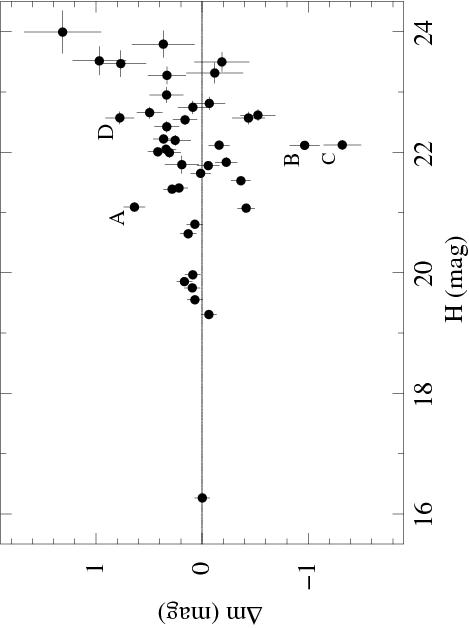}
      \caption{$H-$band magnitude of sources detected within the X--ray error circle (measured on 2010 Aug 30) $vs.$ the magnitude difference between the Aug 30 and Aug 31 epochs. The objects with $\Delta m >0$ mag are brighter in the second epoch. Sources marked as ``ABCD'' show variability larger than the observed dispersion for objects with similar brightness. Errors are at $1\sigma$ c.l., magnitudes are in the Vega system and uncorrected for Galactic extinction. 
	    }
       \label{fig:lc}
   \end{figure}

\begin{figure}
   \centering
   \includegraphics[angle=0,width=\columnwidth]{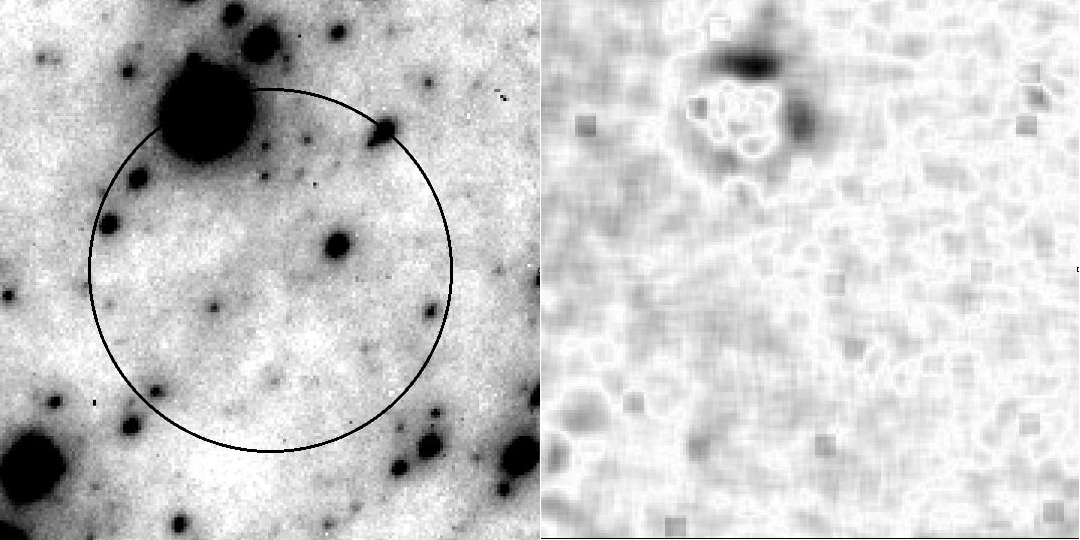}
      \caption{{\it Left panel}: The field of Swift J1749.4--2807 observed in the $H-$band on 2010 Aug 30. The black circle represents the X--ray error circle (1.6'' radius, $90\%$ c.l.). North is up and East is left, image size is $5.4'' \times 5.4''$. {\it Right panel}: The result of image subtraction between the images obtained on Aug 30 and Aug 31. 
	    }
       \label{fig:fc_isis}
   \end{figure}

In order to verify the presence of variable candidates, we carried out PSF-photometry of all the objects visible (above 3$\sigma$ c.l.) in both images which fall within the X--ray error circle (41 sources satisfy these criteria).
We plot in Fig.~\ref{fig:lc} the measured $H-$band magnitudes of epoch 1 vs. epoch 2. As expected, the dispersion increases for fainter objects. Apart from four objects (marked as ``ABCD'') no source is seen to significantly ($\geq 3\sigma$) vary more than 0.09, 0.27, 0.44 and 0.58 magnitudes for objects with $H < 21$, $21 < H < 22$, $22 < H < 23$ and $23 < H < 24$ mag, respectively. 
In order to further investigate the presence of any variable object, we performed image subtraction with the ISIS package (\cite{Al00}; \cite{Al98}) on our two images, using the image obtained on Aug 30 as a reference. The result of the image subtraction analysis is shown in the right panel of Fig.~\ref{fig:fc_isis}. No significant residual is present within the X--ray error circle (excluding the regions close to the brightest star of the field). We note that the background varies considerably between the two epochs, possibly affecting the result of the image subtraction. It is thus possible that such background fluctuations  overcome in the subtracted image the scatter observed with PSF photometry. 
This would imply either that source variability is hidden in the intrinsic dispersion of our photometry or that the NIR counterpart of Swift J1749.4--2807 has $H > 23.5$ mag. 

A list of the candidate NIR counterparts of Swift J1749.4--2807 reporting their positions and $H-$band magnitudes, measured in the two observing epochs, is shown in Table 2.

\section{Discussion and Conclusions}

An important goal of high energy astrophysics is to derive or, at least, place tight constraints on the neutron star equation of state (EOS), probing the properties of ultradense matter. Astrophysics is the only branch of physics that can help since laboratory experiments cannot probe the relevant high-density regime (Lattimer \& Prakash 2006). As discussed in Sec. 1, the eclipsing AMXP Swift J1749.4--2807 would be among the best suited object known so far for dynamical studies aimed at mass measurements, because the radial velocity of the compact object $K_1$ and the system inclination $i$ can be measured with good precision. During quiescence, the optical/NIR flux of a low-mass X--ray transient is dominated by the companion star. The detection of the quiescent optical/NIR counterpart of Swift J1749.4--2807 would therefore make it possible to measure the companion star radial velocity $K_2$ providing the missing information to fully resolve the system and precisely determine its masses. 

The companion star of Swift J1749.4--2807 can, in principle, be a young brown dwarf or a white dwarf, but both are too small to assure the filling of the Roche lobe, for the measured 8.82 hr orbital period. The companion star is therefore a (possibly evolved) main sequence star. From geometrical considerations in an eclipsing system, Markwardt et al. (2010) and Altamirano et al. (2011) derived a companion mass in the 0.46--0.81 $M_{\odot}$ range for a neutron star with mass in the 0.8--2.2 $M_{\odot}$ range, and system inclinations in the $74.4^{\circ}-77.9^{\circ}$ range. Assuming an unirradiated main sequence companion star, these mass limits imply that the spectral type of the companion star should be in the K0V-M0V range. Independently of the compact object mass, the mean density ($<\rho>$) of a Roche lobe-filling companion star can be determined solely by the binary period: $<\rho> = 113{P_h}^{-2}$ g cm$^{-3}$, where $P_h$ is the orbital period in hours (see, e.g., Faulkner, Flannery \& Warner 1972). For Swift J1749.4--2807, with $P_h = 8.82$ hr we obtain $M_2=0.8-1.0$ $M_{\odot}$ for the case of a main sequence companion, i.e. a G2V-K0V type star (Cox 2000). This makes the hypothesis of a main sequence companion star with spectral type later than K less favored.
Alternatively, a lower mass donor star can fill its Roche-lobe if it evolves within a Hubble time. 
According to this scenario, the system must have experienced the evolution of a more massive star before the onset of mass transfer, creating a common envelope phase with unstable mass transfer in a thermal time scale. The final product of this process is a low mass x--ray binary consisting in a companion star with an evolved nucleus trasferring matter towards the compact object (see, e.g., Tauris \& van den Heuvel 2006).
Using the evolutionary tracks of companion stars computed by Schenker \& King (2002), we obtain $M_2 \sim 0.3-0.4$ M$_{\odot}$ for a system like Swift J1749.4--2807 in the limiting case of a donor star with nearly all helium nucleus. 
These considerations are in agreement with the $M_2$ estimates of Markwardt et al. (2010) and Altamirano et al. (2011) based on the system eclipses, with their lower and upper limit corresponding to an evolved or main sequence companion star, respectively.

In light of these considerations, knowing the source distance and the amount of interstellar absorption in its direction it is possible to made an estimate of the expected quiescent NIR luminosity of Swift J1749.4--2807. The detection of a type-I X--ray burst during the 2006 outburst provided an upper limit to the source distance of $6.7 \pm 1.3$ kpc (\cite{Wijn09}). 
The measured X--ray absorbing column density ($N_H = 3.0 \times 10^{22}$ cm$^{-2}$; Ferrigno et al. 2011) is considerably larger than the total Galactic column density in the direction of the source ($N_H = 1.1 \times 10^{22}$ cm$^{-2}$).
This might indicate that a fraction of it is local to the system and possibly ionized (i.e. not contributing to the optical/NIR absorption). Assuming that the full observed column density contributes to the optical/NIR absorption we estimate a color excess $E(B-V) = 5.41$ and $A_H = 2.76$ mag (using the $N_H/E(B-V)$ conversion of \cite{PS95} and the Galactic extinction curves of \cite{Fi99}) and would expect $H-$band apparent magnitudes in the range $20.4-21.3$ for K0V and $21.7-22.6$ for M0V main sequence stars (Cox 2000). These are conservative estimates that do not take into account the fact that the companion star might be subjected to some irradiation from the compact object, even if the system is in quiescence. Furthermore, a lower effective absorption in the NIR band (or an evolved companion star) would result in brighter objects. 
The $3\sigma$ upper limits of our two VLT observations are $H > 23.5$ and $H > 22.8$, respectively. Under these assumptions, we consider realistic the possibility that we have actually detected the NIR counterpart of Swift J1749.4--2807 in our VLT--NACO images. As discussed in Sec. 2, the companion star should be searched primarily among the objects that show variability between the two consecutive night of NACO observations. Assuming a variation of at least 0.1 mag for the light curve of the Swift J1749.4--2807 companion star (as observed in other AMXPs, Shahbaz et al. 1998; Welsh et al. 2000; D'Avanzo et al. 2009 and references therein), we can exclude all objects with $H < 21$ mag (Fig. 1). The remaining objects with $0.4 < \Delta m < 0.6$ are in principle all valid candidates. 
Unfortunately, with just two epochs of observation we cannot discriminate between ``real'' variability and scatter due to the statistical dispersion of our photometry. We just point out that the four sources marked as ``ABCD'', with $H=21.1$, $H=22.1$, $H=22.1$ and $H=22.6$ mag, respectively, (Fig. 2 and Table 2) appear to be promising candidates, because their variability is larger than the observed dispersion for objects with similar magnitudes. 

Finally, we note that the result of our image subtraction (Sec. 2) show a residual near the brightest star within the X--ray error circle (right panel of Fig. 2). If real, this would imply that the NIR quiescent counterpart of Swift J1749.4--2807 is ``hidden'' by a bright ($H=16.3$ mag) interloper (as observed for Aql X-1; \cite{Ch99}) making any future study during quiescence really challenging. However, the variability associated to such residual is not highly significant and can therefore be related to a spurious bad subtraction of the bright star. 

Future phase-resolved NIR observations (performed with large aperture telescopes and adaptive optics) carried out covering a full orbital period would be fundamental to pinpoint the quiescent counterpart of Swift J1749.4--2807 trough the measure of its orbital-modulated variability opening the road for dynamical studies of this very promising source.

\begin{table*}
   \centering
\caption{Candidate NIR counterparts of Swift J1749.4--2807 (see Fig. 2). Magnitudes are in the Vega system and are not corrected for Galactic extinction. Errors are at $1\sigma$ c.l.}
\begin{tabular}{cccc}
\hline
Candidate ID &  Position	      &  $H-$band magnitude   & $H-$band magnitude\\
             &  RA(J2000), Dec(J2000) &   (Aug 30)            &   (Aug 31) 	  \\ 
\hline
 1    & 17:49:31.88, -28:08:03.3     & $16.26 \pm 0.05$      & $16.27 \pm 0.05$   \\
 2    & 17:49:31.79, -28:08:04.5     & $19.31 \pm 0.05$      & $19.37 \pm 0.05$   \\
 3    & 17:49:31.75, -28:08:03.5     & $19.55 \pm 0.05$      & $19.48 \pm 0.05$   \\
 4    & 17:49:31.94, -28:08:04.3     & $19.75 \pm 0.05$      & $19.66 \pm 0.05$   \\
 5    & 17:49:31.92, -28:08:03.9     & $19.85 \pm 0.05$      & $19.69 \pm 0.05$   \\
 6    & 17:49:31.92, -28:08:06.1     & $19.97 \pm 0.05$      & $19.88 \pm 0.05$   \\
 7    & 17:49:31.91, -28:08:05.8     & $20.65 \pm 0.05$      & $20.52 \pm 0.05$   \\
 8    & 17:49:31.72, -28:08:05.1     & $20.81 \pm 0.05$      & $20.74 \pm 0.05$   \\
 9    & 17:49:31.87, -28:08:05.0     & $21.07 \pm 0.06$      & $21.49 \pm 0.06$   \\
10(A) & 17:49:31.84, -28:08:03.6     & $21.09 \pm 0.08$      & $20.45 \pm 0.07$   \\
11    & 17:49:31.81, -28:08:03.5     & $21.39 \pm 0.06$      & $21.10 \pm 0.06$   \\
12    & 17:49:31.83, -28:08:03.9     & $21.41 \pm 0.06$      & $21.19 \pm 0.06$   \\
13    & 17:49:31.74, -28:08:06.1     & $21.53 \pm 0.06$      & $21.89 \pm 0.07$   \\
14    & 17:49:31.95, -28:08:04.0     & $21.65 \pm 0.07$      & $21.64 \pm 0.06$   \\
15    & 17:49:31.82, -28:08:03.0     & $21.78 \pm 0.07$      & $21.84 \pm 0.07$   \\
16    & 17:49:31.96, -28:08:04.9     & $21.79 \pm 0.15$      & $21.60 \pm 0.06$   \\
17    & 17:49:31.76, -28:08:03.6     & $21.83 \pm 0.08$      & $22.06 \pm 0.07$   \\
18    & 17:49:31.81, -28:08:03.2     & $21.99 \pm 0.08$      & $21.69 \pm 0.07$   \\
19    & 17:49:31.81, -28:08:03.9     & $22.01 \pm 0.07$      & $21.59 \pm 0.06$   \\
20    & 17:49:31.83, -28:08:05.7     & $22.05 \pm 0.07$      & $21.71 \pm 0.06$   \\
21(B) & 17:49:31.80, -28:08:04.8     & $22.11 \pm 0.07$      & $23.08 \pm 0.12$   \\
22    & 17:49:31.94, -28:08:05.0     & $22.12 \pm 0.06$      & $22.28 \pm 0.07$   \\
23(C) & 17:49:31.85, -28:08:05.0     & $22.12 \pm 0.08$      & $23.44 \pm 0.16$   \\
24    & 17:49:31.77, -28:08:05.4     & $22.20 \pm 0.08$      & $21.94 \pm 0.12$   \\
25    & 17:49:31.72, -28:08:05.5     & $22.22 \pm 0.07$      & $21.86 \pm 0.07$   \\
26    & 17:49:31.93, -28:08:05.9     & $22.42 \pm 0.09$      & $22.09 \pm 0.07$   \\
27    & 17:49:31.73, -28:08:05.4     & $22.54 \pm 0.08$      & $22.38 \pm 0.08$   \\
28    & 17:49:31.91, -28:08:04.3     & $22.57 \pm 0.10$      & $23.00 \pm 0.12$   \\
29(D) & 17:49:31.79, -28:08:03.8     & $22.57 \pm 0.10$      & $21.79 \pm 0.09$   \\
30    & 17:49:31.76, -28:08:05.6     & $22.62 \pm 0.09$      & $23.14 \pm 0.14$   \\
31    & 17:49:31.86, -28:08:05.9     & $22.66 \pm 0.10$      & $22.16 \pm 0.07$   \\
32    & 17:49:31.73, -28:08:04.5     & $22.75 \pm 0.11$      & $22.66 \pm 0.09$   \\
33    & 17:49:31.81, -28:08:04.3     & $22.81 \pm 0.11$      & $22.88 \pm 0.11$   \\
34    & 17:49:31.93, -28:08:05.3     & $22.95 \pm 0.13$      & $22.61 \pm 0.09$   \\
35    & 17:49:31.78, -28:08:05.9     & $23.28 \pm 0.14$      & $22.94 \pm 0.11$   \\
36    & 17:49:31.96, -28:08:04.7     & $23.31 \pm 0.17$      & $23.43 \pm 0.20$   \\
37    & 17:49:31.83, -28:08:05.6     & $23.47 \pm 0.22$      & $22.70 \pm 0.10$   \\
38    & 17:49:31.82, -28:08:06.4     & $23.50 \pm 0.16$      & $23.68 \pm 0.20$   \\
39    & 17:49:31.82, -28:08:05.1     & $23.52 \pm 0.24$      & $22.55 \pm 0.08$   \\
40    & 17:49:31.74, -28:08:04.4     & $23.79 \pm 0.22$      & $23.43 \pm 0.20$   \\
41    & 17:49:31.76, -28:08:05.4     & $23.99 \pm 0.35$      & $22.68 \pm 0.09$   \\
\hline
\end{tabular}
\label{tab:candidates}
\end{table*}

\begin{acknowledgements}
We thank the director of the European Southern Observatory for granting Director's Discretionary Time (ID 285.D-5030(A)). PDA and SC acknowledge tha Italian Space Agency for financial supportthrough the project ASI I/009/10/0. The research leading to these results has received funding from the European Community's Seventh Framework Programme (FP7/2007-2013) under grant agreement number ITN 215212 ``Black Hole Universe'' and from the spanish MEC under the Consolider-Ingenio 2010 Program grant CSD2006-00070: First Science with the GTC (http://www.iac.es/consolider-ingenio-gtc/).
\end{acknowledgements}


\begin{thebibliography}{}

\bibitem[Alard 2000]{Al00}
Alard, C. 2000, A\&AS, 144, 363

\bibitem[Alard \& Lupton 1998]{Al98}
Alard, C., Lupton, R. 1998, ApJ, 503, 325

\bibitem[Altamirano et al. 2008]{Al08}
Altamirano, D., Casella, P., Patruno, A., Wijnands, R., van der Klis, M. 2008, ApJ, 674, L45

\bibitem[Altamirano et al. 2010]{Al10}
Altamirano, D., Patruno, A., Heinke, C. O., et al, 2010, ApJ, 712, L58

\bibitem[Altamirano et al. 2011]{Al11}
Altamirano, D., Cavecchi, Y., Patruno, A., et al. 2011, ApJ, 727, L18

\bibitem[Blustin et al. 2006]{Blu06}
Blustin, A. J., Schady, P., Pandey, S. B. 2006, GCN Circ, 5207

\bibitem[Campana et al. 2003]{Ca03}
Campana, S., Ravasio, M., Israel, G. L., Mangano, V., Belloni, T. 2003, ApJ, 594, L39

\bibitem[Campana 2009]{Ca09}
Campana, S. 2009, ApJ, 699, 1144

\bibitem[Casella et al. 2008]{Ca08}
Casella, P., Altamirano, D., Patruno, A., Wijnands, R., van der Klis, M. 2008, ApJ, 674, L41

\bibitem[Chakrabarty \& Morgan 1998]{CM98}
Chakrabarty, D., Morgan, E. H. 1998, Nature, 394, 346

\bibitem[Chakrabarty, Jonker \& Markwardt 2010]{CJM10}
Chakrabarty, D., Jonker, P. G. and Markwardt, C. B. 2010, ATel 2585

\bibitem[Chevalier et al. 1999]{Ch99}
Chevalier, C., Ilovaisky, S. A., Leisy, P., Patat, F. 1999, A\&A, 347, L51

\bibitem[Cox 2000]{Co00}
Cox, A. N., Allen's astrophysical quantities, 4th ed. Publisher: New York: AIP Press; Springer, 2000

\bibitem[D'Avanzo et al. 2009]{PDA09}
D'Avanzo, P., Campana, S., Casares, J., et al. 2009, A\&A, 508, 297

\bibitem[Falanga et al. 2005]{Fa05}
Falanga, M., Kuiper, L., Poutanen, J., et al. 2005, A\&A, 444, 15

\bibitem[Faulkner, Flannery \& Warner 1972]{FFW72}
Faulkner, J., Flannery, B. P., Warner, B. 1972, ApJ, 175, L79

\bibitem[Ferrigno et al. 2011]{Fe11}
Ferrigno, C., Bozzo, E., Falanga, M., et al. 2011, A\&A, 525, 48

\bibitem[Fitzpatrick 1999]{Fi99}
Fitzpatrick, E. L. 1999, PASP, 111, 63

\bibitem[Galloway et al. 2002]{Ga02}
Galloway, D. K., Chakrabarty, D., Morgan, E. H., Remillard, R. A. 2002, ApJ, 576, L137

\bibitem[Galloway et al. 2005]{Ga05}
Galloway, D. K., Markwardt, C. B., Morgan, E. H., Chakrabarty, D., Strohmayer, T. E. 2005, ApJ, 622, L45

\bibitem[Hynes 2010]{Hy10}
Hynes, R. 2010, eprint arXiv:1010.5770

\bibitem[Kaaret et al. 2006]{Ka06}
Kaaret, P.; Morgan, E. H.; Vanderspek, R.; Tomsick, J. A. 2006, ApJ, 638, 963

\bibitem[Khamitov et al. 2010]{Kha06}
Khamitov, I., Bikmaev, I., Sakhibullin, N., et al.  2006, GCN Circ, 5205

\bibitem[Krimm et al. 2007]{Kr07}
Krimm, H. A., Markwardt, C. B., Deloye, C. J., et al. 2007, ApJ, 668, L147

\bibitem[Kubanek et al. 2006]{Ku06}
Kubanek, P., Jelinek, M., French, J. 2006, GCN Circ, 5199

\bibitem[Lattimer \& Prakash 2006]{LP06}
Lattimer, J. M., Prakash, M. 2006, NuPhA, 777, 479

\bibitem[Markwardt et al. 2002]{Ma02}
Markwardt, C. B., Swank, J. H., Strohmayer, T. E., in't Zand, J. J. M., Marshall, F. E. 2002, ApJ, 575, L21

\bibitem[Markwardt \& Strohmayer 2010]{Ma10}
Markwardt, C. B., Strohmayer, T. E. 2010, ApJ, 717, L149

\bibitem[Melandri et al. 2006]{Me06}
Melandri, A., Di Stefano, E., Covino, S., et al. 2006, GCN Circ., 5229

\bibitem[Papitto et al. 2010]{Pa10}
Papitto, A., Riggio, A., di Salvo, T., et al. 2010, MNRAS, 407, 2575

\bibitem[Papitto et al. 2011]{Pa11}
Papitto, A., Ferrigno, C.,m Bozzo, E., et al. 2011, ATel 3556 

\bibitem[Predehl \& Schmitt 1995]{PS95}
Predehl, P., Schmitt, J. H. M. M. 1995, A\&A, 293, 889

\bibitem[Schenker \& King 2002]{SK2002}
Schenker, K., King, A. R. 2002, in Gansicke B. T., Bevermann K., Reinsch
K., eds, ASP Conf. Vol. 261, The Physics of Cataclysmic Varibles and
Related Objects. Astron Soc. Pac., San Francisco, p. 242

\bibitem[Shahbaz et al. 1998]{Sh98}
Shahbaz, T., Thorstensen, J. R., Charles, P. A., Sherman, N. D. 1998, MNRAS, 296, 1004

\bibitem[Steeghs \& Casares 2002]{SC02}
Steeghs, D., Casares, J. 2002, ApJ, 568, 273

\bibitem[Strohmayer et al. 2003]{Str03}
Strohmayer, T. E., Markwardt, C. B., Swank, J. H., in't Zand, J. 2003, ApJ, 596, L67

\bibitem[Tauris \& van den Heuvel 2006]{TVDH06}
Tauris, T. M. and  van den Heuvel, E. P. J. 2006, in Compact Stellar X-Ray Sources, eds. W.H.G. Lewin and M. van der Klis, Cambridge University Press 

\bibitem[Thorsett \& Chakrabarty 1999]{TC99}
Thorsett, S. E., Chakrabarty, D. 1999, ApJ, 512,288

\bibitem[Yang et al. 2010]{Ya10}
Yang, Y. J., Russell, R., Wijnands, R., et al. 2010, ATel, 2585

\bibitem[Welsh et al. 2000]{We05}
Welsh, W. F., Robinson, E. L., Young, P. 2000, AJ, 120, 943	

\bibitem[Wijnands \& van der Klis 1998]{WV98}
Wijnands, R., van der Klis, M. 1998, Nature, 394, 344

\bibitem[Wijnands 2005]{Wijn05}
Wijnands, R.  2005, arXiv:astro-ph/0501264v1

\bibitem[Wijnands et al. 2009]{Wijn09}
Wijnands, R., Rol, E., Cackett, E., et al. 2009, MNRAS, 393, 126

\end{thebibliography}

\end{document}